\documentclass[twocolumn, 10pt,floatfix]{article}

\usepackage{graphicx}
\usepackage{amsmath,amssymb,amsfonts}
\usepackage{color}
\usepackage{hyperref}
\usepackage{textcomp}
\usepackage{verbatim}
\usepackage[numbers]{natbib}
\usepackage[utf8]{inputenc}

\author{Miguel V. Vitorino, Arthur Vieira, M\'{a}rio S Rodrigues \\
\small{1- Biosystems $\&$ Integrative Sciences Institute, Faculdade de Ciências, Universidade de Lisboa, 1749-016 Lisboa, Portugal} \\
\small{2- Departamento de Física, Faculdade de Ciências, Universidade de Lisboa, 1749-016 Lisboa, Portugal} \\
\begin{minipage}{430pt}\small{
\vspace{12pt}
Sliding friction is ubiquitous in nature as are harmonic oscillators. 
However, when treating harmonic oscillators the effect of sliding friction is often neglected. 
Here, we propose a simple analytical model to include both viscous and sliding friction 
in common harmonic oscillator equations, allowing to separate these different types of dissipation. 
To compare this model with experimental data, 
a nanometric vibration was imposed on a quartz tuning fork, 
while an atomic force microscope tip was used to disturb its motion. 
We analyzed tuning fork resonance curves and `ring down' experiments and for each case calculated
the amount of sliding friction and of viscous damping, finding an agreement between the two different experiments and the model proposed.}
\end{minipage}
}

\title{Effect of sliding friction in harmonic oscillators.}

\begin{document}

\maketitle

\let\thefootnote\relax\footnote{$^*$ Corresponding author: \textit{mmrodrigues@fc.ul.pt}}

Friction is an ubiquitous force that manifests from macro to nanoscales and is present almost everywhere in nature, in a wide variety of applications and extremely different industries, including automotive manufacturing, winter sports gear and nanotechnology, where new devices present a very high surface-to-volume ratio and a strong influence by surface forces. 
The study of friction has played an important role in the scientific community for centuries, 
from the formulation of the Amontons-Coulomb laws more than 300 years ago\cite{amontons1699resistance}, until the more recent studies of the origin of these forces at atomic scales. However, despite its universal character, friction forces still represent a challenge in the quantitative analysis of many systems. For example, its effect in the motion of an harmonic oscillators is not trivial. Despite several different approaches, from numerical ones\cite{csernak2006periodic} to heuristic arguments\cite{marchewka2004oscillator}, an analytical simple solution to this problem is still lacking as demonstrated by the amount of propositions to solve it in current research\cite{hong2000coulomb,xia2003modelling,korman2014harmonic,lima2015stick}. 
Perhaps more notoriously this complication arises in the field of tribology. 

Since the 1970s tribology has benefited from the invention of a number of new tools, 
such as the atomic force microscope (AFM)\cite{binnig1986afm,mate1987atomic}, 
or the quartz crystal microbalance (QCM)\cite{krim1988damping,krim2012friction}, 
that have allowed for an unprecedented advance in the mechanistic explanation of the different friction 
regimes\cite{krim1991nanotribology,Yoshizawa1993fundamental,yoshizawa1993fundamentaloi,riedo2003interaction,tambe2005friction,urbakh2010nanotribology,chiu2012adhesion}. 
However, friction itself is still not completely understood, 
and a number of authors have worked to combine the advantages of AFM and QCM in an effort to generalize nanotribology models
\cite{borovsky2001measuring,berg2003high,inoue2012dynamical,nigues2014ultrahigh,krim2012friction}. 
Despite promising results, these approaches currently lack the theoretical insight allowing the analysis 
of a system in which there is sliding friction independent of the sliding speed magnitude, or a more general non-linear friction law. 
Some progress has been done addressing non linear friction forces\cite{berg2003nonlinear}, 
but the applicability of these results is often limited to the particular force law considered. 
In general a simple viscous law is assumed, which may or may not be applicable, 
depending on the physical system under analysis.

In this letter we present analytical solutions to the common harmonic oscillator equations when a sliding friction force term is added. To demonstrate the validity of these solutions nanotribology experiments were performed, using an AFM to disturb the movement of a quartz Tuning Fork (TF), and this disturbance was compared to our model.

We focus on the effect of friction in oscillators that can still swing back and forth a few times if released from rest. 
Roughly speaking, if $F_f$ is the friction force, $A_i$ the initial displacement 
and $k$, $m$, $\gamma$ the oscillator spring constant, mass and damping coefficient respectively, this work focuses on the case in which $4 F_f < \pi k A_i$ and the quality factor $Q>2$ with $Q\equiv\sqrt{km}/\gamma=k/\gamma \omega_0$.

\begin{figure}[tb]
\centering
\includegraphics[]{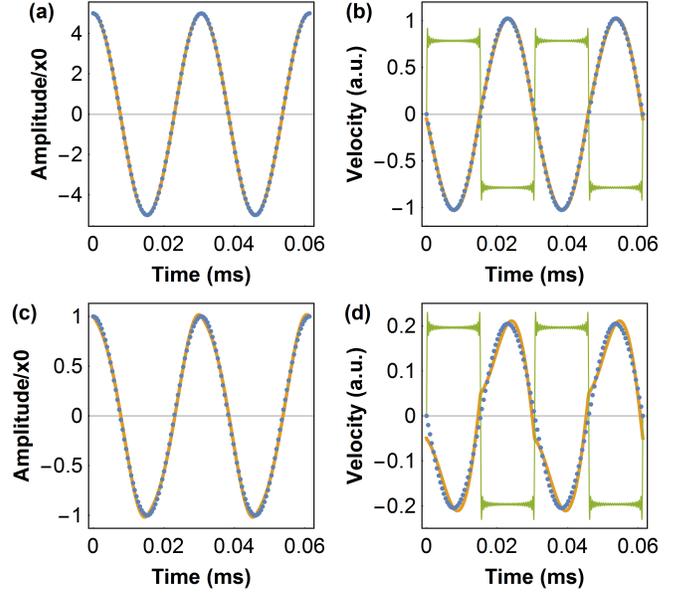}
\caption{Oscillator amplitude (a) and velocity (b) versus time
for a system submitted to a friction such that the oscillator gain is $G=5$, 
showing the leading (blue, points) and the first 30 terms of the Fourier series (orange, continuous);
in (a) the two curves are indistinguishable, in (b) 
we additionally plot the friction force (green), rescaled for clarity; 
(c) and (d) the same as in (a) and (b) for a higher friction force such that  $G=1$
illustrating conditions for which the model starts loosing accuracy.  
}
\label{fig:1}
\end{figure}

We begin by considering steady state motion and we assume the oscillator is moving with periodic, however not simple harmonic motion due to the presence of sliding friction. 
Additionally, we assume the system inverts its velocity at multiples of its period $T$.
As a consequence, friction is a square wave with period $T$, for which the phase can arbitrarily be chosen as $\phi_{friction}=0$. 
If the magnitude of the friction force is $F_f$ then its Fourier series can be written as:
\begin{equation}
  F_f(\omega)=\frac{4 F_f}{n \pi} \sin(n \omega t),
  \label{eq:1}
\end{equation}
\begin{flushright} with n=1,3,5... \end{flushright}
\noindent where the summation over $n$ is implied. Therefore, in the case where the oscillator period equals the excitation force period, the equation of motion will be:
\begin{equation}
  m \ddot{x}=-k(x-x_0 \cos(\omega t-\phi_0))-\gamma \dot{x}-\frac{4 F_f}{n \pi} \sin(n \omega t) 
  \label{eq:3}
\end{equation}
\begin{flushright} with n=1,3,5... \end{flushright}
where $x_0$ is the excitation amplitude and $\phi_0$ the excitation phase, yet to be determined. The steady state solution for the equation above is:
\begin{equation}
 x(t)= R \cos(\omega t-\phi_1)+ R_n \cos(n \omega t-\phi_n)
\label{eq:4}
 \end{equation}
 \begin{flushright} with n=3,5,7,... \end{flushright}
Replacing this solution in Eq. \ref{eq:3} and solving for the
Fourier components $R_n$ and $\phi_n$, we find:
\begin{equation}
R_n=\frac{4 F_f/\pi n}{\sqrt{(k-m (n w)^2)^2+ \gamma^2 (n \omega)^2}}
\label{eq:Rn}
\end{equation}
\begin{equation}
 \phi_n=\arctan\left(\frac{n \gamma \omega}{k-m n^2 \omega^2} \right)
\end{equation}
Before solving for the leading term one must make a note. If we put $\phi_1=0$ in Eq. \ref{eq:4}, this equation is still a solution of Eq. \ref{eq:3}, and indeed it is the solution we will explore later. However, such situation does not guarantee that the oscillator changes the signal of the speed in phase with the friction force. One can differentiate Eq. \ref{eq:4} to find the speed, and impose zero speed each time the friction force is zero, i.e. when $\omega t=m \pi$, with $m$ integer, leading to:
\begin{equation}
 R \sin(\phi_1)=-n R_n \sin(\phi_n)
\end{equation}
\noindent where again summation over $n$ is implied, and:
\begin{equation}
 \phi_1=\arcsin\left( \sum_{n=3,5}^{\infty} \frac{4 F_f n \gamma \omega/R \pi}{(k-m (n w)^2)^2+ \gamma^2 (n \omega)^2} \right) 
\end{equation}
Within the limits mentioned before one can show that $\phi_1 \approx 0$. Assuming this is true, the problem simplifies substantially (see Fig. \ref{fig:1}), the solution for the leading term in Eq. \ref{eq:3} becoming:
\begin{equation}
R=\frac{-4F_f \gamma \omega+\sqrt{\pi^2 k^2 x_0^2 Z^2-16 F_f^2 (k-m \omega^2)^2}  }{\pi Z^2}
\label{eq:leading}
\end{equation}
\noindent with $Z=\sqrt{(k-m \omega^2)^2+\gamma^2 \omega^2}$. 
For excitation frequencies close to the resonance frequency, 
the oscillator filters out the higher components (n=3,5,...) of the Fourier series.
The excitation frequency appears multiplied by $n$ in the denominator of Eq. \ref{eq:Rn},
hence, the contribution of the higher terms is negligible.
Consequently, the oscillator amplitude of vibration is effectively described by Eq. \ref{eq:leading}. 
We propose below a simplified version of this equation. 

The oscillator at resonance frequency has a gain $G=R_0/x_0$, which in absence of sliding friction
is simply the quality factor $Q$.
Fig. \ref{fig:1} presents amplitude and 
velocity calculated with only the first and 
with 30 terms of the Fourier series, in two limiting conditions: 
starting with an oscillator with a gain $G=10^4$ ($=Q$), friction is increased such that the oscillator gain becomes $G=5$ (Fig. \ref{fig:1}(a) and (b)) 
and $G=1$ (Fig. \ref{fig:1}(c) and (d)). 
These cases correspond to friction forces such that $4 F_f \approx \pi k x_0$.
Fig. \ref{fig:1} illustrates that even for overdamping friction forces and very small oscillator gains $G$,
Eq. \ref{eq:leading}, which neglects the higher order terms of the Fourier series, is very effective at describing the system.
For very low vibration amplitudes, $R$, smaller than the excitation amplitude $x_0$, 
this model looses accuracy because the friction force is no longer in phase with velocity, 
$\phi_1\approx 0$ becoming a weaker approximation. 
Additionally, for such extreme cases one would certainly need 
to consider also static friction i.e when the oscillator speed is zero, 
as it may rest at zero speed for a time longer than in the harmonic situation. 
In fact, we suggest that the effect of static friction in harmonic oscillators 
can also be treated using an appropriate Fourier series.

Eq. \ref{eq:leading} is generally difficult to fit to experimental data. 
However this equation looks like and can be approximated to a Lorentzian and this fact may be deceiving, for one may erroneously conclude all forces involved are linear. Nonetheless it is useful to express it in terms of a Lorentzian dependence which is routinely fitted to resonance curves.

From Eq. \ref{eq:leading} it follows that at resonance the amplitude $R_0$ is given by:
\begin{equation}
 R_0 = Q \left(x_0-\frac{4 F_f}{\pi k}\right) = x_f Q_f
 \label{eq:xfqf}
\end{equation}
\noindent where $x_f$ and $Q_f$ are defined as the effective parameters obtained from a Lorentzian fit, 
$Q$ being the quality factor defined such that it depends on damping but not on sliding friction. 
Similarly, for a given amplitude $R$ one can define an effective damping $\gamma_f$, 
by adding the leading term of the sliding friction series to the 
term containing the damping coefficient $\gamma$ in Eq. \ref{eq:3}:
\begin{equation}
 \gamma_f=\gamma+\frac{4 F_f}{\pi \omega_0 R}
 \label{eq:g_f}
\end{equation}
The amplitude $R$ must be proportional to $x_0$ and to the effective quality factor 
$Q_f\equiv k/(\gamma_f \omega_0)$; 
thus putting $R = a x_0 k/(\gamma_f \omega_0)$, with $a$ a proportionality constant, and 
solving for Eqs. \ref{eq:xfqf} and \ref{eq:g_f} one finds:
\begin{equation}
\gamma_f=\gamma (1-4 F_f/a \pi k x_0)^{-1}
\label{eq:gamaf}
\end{equation}
\noindent or equivalently,
\begin{equation}
Q_f= Q (1-4 F_f/a \pi k x_0)
\label{eq:gamaf2}
\end{equation}
\noindent and 
\begin{equation}
x_f=x_0 \frac{1-4 F_f/\pi k x_0}{1-4 F_f/a k \pi x_0}
\label{eq:xf}
\end{equation}
Consequently, the solution presented in Eq. \ref{eq:leading} can be written in the usual form:
\begin{equation}
R=\frac{x_f \omega_0^2}{\sqrt{\left(\omega_0^2-\omega^2\right)^2 + \left(\frac{\omega \omega_0}{Q_f}\right)^2}}
\label{eq:finalLorenz}
\end{equation}
Finally, comparing Eqs. \ref{eq:finalLorenz} with \ref{eq:leading} yields $a \approx 4/3$, allowing us to calculate $x_f$ and $Q_f$.

We turn now to a common situation where the oscillator is left to stop after released from an initial amplitude, $A_i$. 
From the previous analysis one concludes that the resonance frequency is not affected by pure sliding friction forces.
Additionally we know that an harmonic oscillator vibrates at its resonance if released from rest. 
Consequently, as before, we consider the friction force as a square wave with 
frequency equal to the resonance frequency $\omega_r$. 
The situation can be described by the following equation:
\begin{equation}
  m \ddot{x}+\gamma \dot{x}+k x=\frac{4 F_f}{\pi} \sin(\omega_r t)
 \label{eq:RD1}
\end{equation}
\noindent where the higher order terms have been neglected for reasons explained earlier. 
Eq. \ref{eq:RD1} must be solved with initial conditions $x(0)=A_i$. 
Obviously, this equation only describes the system until its amplitude is zero, as for 
instants after that the equation above makes no sense. 
The solution resulting from considering $4 Q^2>>1$, is: 
\begin{equation}
 x(t)=\left[\frac{-4 F_f }{\pi \gamma w_0}+\left(A_i+\frac{4 F_f }{\pi \gamma w_0}\right) e^{-\gamma t/2 m}\right]\cos(\omega_r t)
 \label{eq:RD2}
\end{equation}
Unlike the common harmonic oscillator, 
the amplitude decay is not just a simple exponential, 
and an `offset' appears as a signature of sliding friction. 
This equation provides a very simple way to test if the system is subject to sliding friction since in that case, 
and unlike when it is not, $\textrm{Log}[v(t)]$ does not give a straight line. 

The same expression can be deduced in a much simpler way. 
Considering the system with an initial kinetic energy $1/2 \ m v_i^2$ and at the equilibrium position, after a certain time the oscillator has gone forth and returned to the initial position, losing an energy $2 F_f A$ due to sliding friction and an energy $(\pi/2) \gamma A v$ due to damping. We can approximate $v$ such that $v(t)=v\sin(\omega_r t)$ between instant $i$ and $f$. 
Thus, during the time it goes back and forth, the oscillator loses energy according to:
\begin{equation}
 \frac{1}{2} m (v_i^2-v_f^2)= 2 F_f \frac{v}{\omega_r}+\frac{\pi}{2} \gamma v^2
 \label{eq:energy}
\end{equation}
Since we are considering a small time interval $dt=\pi/\omega_r$, 
one can approximate $v_i=v+dv/2$ and $v_f=v-dv_m/2$. 
To calculate the rate at which the system is losing energy one must divide Eq. \ref{eq:energy} 
by the time $dt$ it takes the system to go back and forth. 
Eq. \ref{eq:energy} becomes:
\begin{equation}
 -m \frac{dv}{dt}= \frac{2 F_f}{\pi} + \frac{\gamma v}{2}
 \label{eq:energy2}
\end{equation}
\noindent The solution to this differential equation when solved with initial conditions $v=(A_i w_r)$ is identical to that of Eq. \ref{eq:RD2}. The method presented here can also be used to calculate the effect of friction in oscillators for which the damping is proportional to the square of the speed. For that one needs only to recalculate the energy lost due to damping and replace it in Eq. \ref{eq:energy}.

\begin{figure}[tb]
\centering
\includegraphics[]{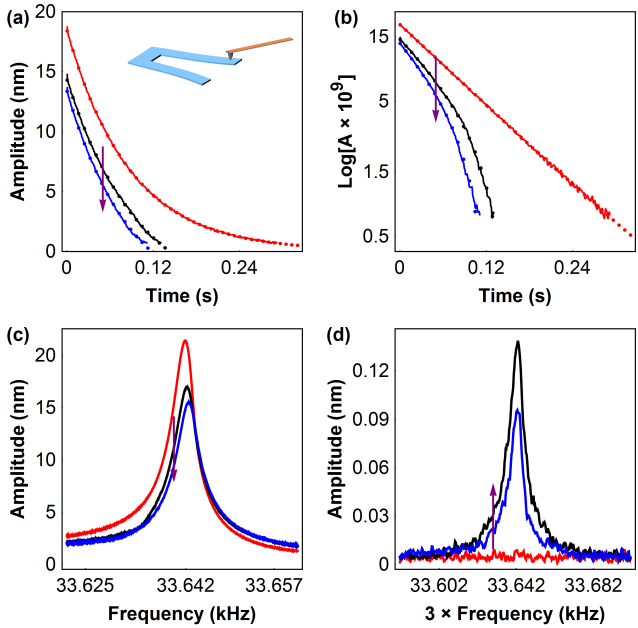}
\caption{
Driving a quartz tuning fork while applying a normal load (0-10 nN): 
(a) and (b) ring down experiments, where the excitation is turned off and the oscillator speed decays. 
Experimental data is presented as solid lines and fits of Eq. \ref{eq:RD2} are presented as dots. 
(c) frequency response of the oscillator with different applied loads. 
(d) frequency response of the oscillator measured around $\omega_0$, 
when driving it near $\omega_0/3$. 
All experiments were performed using similar loads. The red curve corresponds to the free oscillator and the arrow represents increasing load.
}
\label{fig:2}
\end{figure}

To experimentally verify the validity of this model,  
an experimental setup predominant 
in the study of friction with oscillators \cite{borovsky2001measuring,berg2003high,inoue2012dynamical,nigues2014ultrahigh} was used. 
This setup consists of a nanosized AFM tip that exerts force on an harmonic oscillator, 
in this case a quartz tunning fork (TF). 
The symmetric vibration mode of the TF is mechanically excited with a piezoelectric dither and the TF oscillation is measured while the AFM tip applies a constant force on one of its prongs. 
The system is then defined as a silicon tip sliding on a bare quartz surface.
A sketch of the setup can be seen on the inset of Fig. \ref{fig:2}(a). 

The most straightforward way to test the assumption of a viscous friction force law is to perform `ring down' experiments as described previously.
The TF was excited such that its free amplitude of oscillation at $\omega_0$ was close to 25nm. 
A controlled normal load (on the order of a few nN) was then applied by the AFM tip, 
and the excitation was subsequently turned off, 
the amplitude of oscillation being measured until the oscillator stopped. 
Fig. \ref{fig:2}(a) and (b), present the results of this experiment for the free oscillator (red curve) 
and for two different applied loads (black and blue).  
In the unperturbed oscillator, the decay is exponential, evidenced by the linear evolution of Log[$\dot{x}$] with time. 
This indicates a viscous damping of the TF, as expected. 
However, for non-zero applied loads as small as 5 nN, 
the oscillator yields a different ring down decay, 
as seen by the black and blue curves in these figures. 
Experimental data was fitted using Eq. \ref{eq:RD2}, 
and the resulting fits are also plotted on the figures (dots). 
An excellent agreement between experiment and this model was found down to very small oscillating speeds, 
where the tip may be sticking to the TF and the model is not valid anymore.
Additionally we measured the frequency response of the oscillator at different applied loads. 
Fig. \ref{fig:2}(c) presents the results of this experiment, 
where the excitation frequency was scanned around $\omega_0$ and the amplitude of oscillation recorded, 
while controlling the normal load. 
As evidenced in this figure, 
all curves can be fitted quite well by a Lorentzian dependence. 
From the fits of these curves one can extract $x_f$ and $\gamma_f$ and compute $F_f$ and $\gamma$ from Eqs. \ref{eq:gamaf} and \ref{eq:xf}. The results of these fittings from different ring down and resonance curve experiments, at similar loads, can be found in Table \ref{table}.
\begin{table}
\centering
\begin{tabular}{| p{1cm} | p{1cm} | p{1cm} | p{1.65cm} | p{1.65cm} |}
\hline
\centering ${\raise.17ex\hbox{$\scriptstyle\mathtt{\sim}$}}$Load (nN)& \centering $F_{f,rd}$ (nN) & \centering $F_{f,res}$ (nN) & \centering $\gamma_{rd}$ ($\mu \textrm{N} \textrm{m}^{-1} \textrm{s}$) & \centering $\gamma_{res}$ ($\mu \textrm{N} \textrm{m}^{-1} \textrm{s}$)  \tabularnewline
\hline
\centering Free & \centering 0 & \centering 0 & \centering 22 & \centering 20  \tabularnewline
\centering 9 & \centering 15 & \centering 16 & \centering 20 & \centering 18 \tabularnewline
\centering 13 & \centering 17 & \centering 16 & \centering 23 & \centering 20 \tabularnewline
\hline
\end{tabular}
\caption{Fitting parameters extracted from several ring down and transfer function experiments, with different normal loads applied to the oscillator. Suffix `rd' represents ` from ring down curves' and `res' represents from `resonance curves'.}
\label{table}
\end{table}
The fitting parameters from either ring down experiments and resonance curve measurements 
demonstrate that the two experiments are compatible and agree well in describing 
the friction force the oscillator is experiencing. 
Furthermore, and perhaps more importantly, 
these independent measurements demonstrate that there is no significant change in the damping coefficient $\gamma$,
indicating sliding friction as the main mechanism of energy loss.

In fact, we have performed a number of experiments using both methods, with different TFs, cantilevers and loads, and no significant increase of the damping coefficient due to the interaction was detected. For systems akin to ours, the effect of a constant friction force cannot be ignored, even in cases where the oscillator transfer function can be fitted as a Lorentzian. Accurate quantitative results can only be obtained if one includes sliding friction.

Looking at Eq. \ref{eq:Rn}, it becomes quite tempting to conclude that the oscillator can be very effectively excited by driving it at odd subharmonics of the resonance frequency. The oscillator will work out to filter the terms that do not excite it at its resonance frequency. Experimentally this works remarkably well particularly for an excitation frequency of $\omega_0/3$. Fig. \ref{fig:2} shows the response of the oscillator around the resonance frequency $\omega_0$ but excited at $\omega_0/3$ 
in the absence of friction and for two different loads.
Note that in this case the equations developed here do not necessarily apply because our initial hypothesis that the friction frequency was equal to the excitation frequency is not valid. Nonetheless, the fact that the oscillator can be excited at $f=f_0/3$ provides remarkable evidence that the oscillator experiences friction as series of harmonic terms.

In conclusion we have provided
a quantitative method to simultaneously extract the damping and the often forgotten sliding friction 
on experiments involving harmonic oscillators. 
We have developed two independent analytical methods 
to analyze the effect of these dissipation mechanisms on the movement of an oscillator such as a TF, 
and compared our predictions with experiment using an AFM tip and a TF, 
in a common experimental configuration often used in nanotribology. 
This model in conjunction with experimental results demonstrate that
it is not realistic to assume a simple viscous friction law based solely 
on the fact that the oscillator response is of the Lorentzian type. 
Ring down decay experiments exhibit clearly the presence of sliding friction even 
when in the same conditions the resonance curve is rather Lorentzian like. 
We stress, however, that this model is not limited to nanoscale friction, 
but can be applied to even the most simple experiments performed in a physics class, 
providing a simple and analytical answer to a common problem pervasive in nature,
where friction and oscillatory movement are often present but overlooked.


\begin{thebibliography}{23}%
\makeatletter
\providecommand \@ifxundefined [1]{%
 \@ifx{#1\undefined}
}%
\providecommand \@ifnum [1]{%
 \ifnum #1\expandafter \@firstoftwo
 \else \expandafter \@secondoftwo
 \fi
}%
\providecommand \@ifx [1]{%
 \ifx #1\expandafter \@firstoftwo
 \else \expandafter \@secondoftwo
 \fi
}%
\providecommand \natexlab [1]{#1}%
\providecommand \enquote  [1]{``#1''}%
\providecommand \bibnamefont  [1]{#1}%
\providecommand \bibfnamefont [1]{#1}%
\providecommand \citenamefont [1]{#1}%
\providecommand \href@noop [0]{\@secondoftwo}%
\providecommand \href [0]{\begingroup \@sanitize@url \@href}%
\providecommand \@href[1]{\@@startlink{#1}\@@href}%
\providecommand \@@href[1]{\endgroup#1\@@endlink}%
\providecommand \@sanitize@url [0]{\catcode `\\12\catcode `\$12\catcode
  `\&12\catcode `\#12\catcode `\^12\catcode `\_12\catcode `\%12\relax}%
\providecommand \@@startlink[1]{}%
\providecommand \@@endlink[0]{}%
\providecommand \url  [0]{\begingroup\@sanitize@url \@url }%
\providecommand \@url [1]{\endgroup\@href {#1}{\urlprefix }}%
\providecommand \urlprefix  [0]{URL }%
\providecommand \Eprint [0]{\href }%
\providecommand \doibase [0]{http://dx.doi.org/}%
\providecommand \selectlanguage [0]{\@gobble}%
\providecommand \bibinfo  [0]{\@secondoftwo}%
\providecommand \bibfield  [0]{\@secondoftwo}%
\providecommand \translation [1]{[#1]}%
\providecommand \BibitemOpen [0]{}%
\providecommand \bibitemStop [0]{}%
\providecommand \bibitemNoStop [0]{.\EOS\space}%
\providecommand \EOS [0]{\spacefactor3000\relax}%
\providecommand \BibitemShut  [1]{\csname bibitem#1\endcsname}%
\let\auto@bib@innerbib\@empty
\bibitem [{\citenamefont {Amontons}(1699)}]{amontons1699resistance}%
  \BibitemOpen
  \bibfield  {author} {\bibinfo {author} {\bibfnamefont {G.}~\bibnamefont
  {Amontons}},\ }\href@noop {} {\bibfield  {journal} {\bibinfo  {journal} {Mem.
  l'Academie R.}\ } (\bibinfo {year} {1699})}\BibitemShut {NoStop}%
\bibitem [{\citenamefont {Csern{\'a}k}\ and\ \citenamefont
  {St{\'e}p{\'a}n}(2006)}]{csernak2006periodic}%
  \BibitemOpen
  \bibfield  {author} {\bibinfo {author} {\bibfnamefont {G.}~\bibnamefont
  {Csern{\'a}k}}\ and\ \bibinfo {author} {\bibfnamefont {G.}~\bibnamefont
  {St{\'e}p{\'a}n}},\ }\href@noop {} {\bibfield  {journal} {\bibinfo  {journal}
  {Journal of Sound and Vibration}\ }\textbf {\bibinfo {volume} {295}},\
  \bibinfo {pages} {649} (\bibinfo {year} {2006})}\BibitemShut {NoStop}%
\bibitem [{\citenamefont {Marchewka}\ \emph {et~al.}(2004)\citenamefont
  {Marchewka}, \citenamefont {Abbott},\ and\ \citenamefont
  {Beichner}}]{marchewka2004oscillator}%
  \BibitemOpen
  \bibfield  {author} {\bibinfo {author} {\bibfnamefont {A.}~\bibnamefont
  {Marchewka}}, \bibinfo {author} {\bibfnamefont {D.~S.}\ \bibnamefont
  {Abbott}}, \ and\ \bibinfo {author} {\bibfnamefont {R.~J.}\ \bibnamefont
  {Beichner}},\ }\href@noop {} {\bibfield  {journal} {\bibinfo  {journal}
  {American journal of physics}\ }\textbf {\bibinfo {volume} {72}},\ \bibinfo
  {pages} {477} (\bibinfo {year} {2004})}\BibitemShut {NoStop}%
\bibitem [{\citenamefont {Hong}\ and\ \citenamefont
  {Liu}(2000)}]{hong2000coulomb}%
  \BibitemOpen
  \bibfield  {author} {\bibinfo {author} {\bibfnamefont {H.-K.}\ \bibnamefont
  {Hong}}\ and\ \bibinfo {author} {\bibfnamefont {C.-S.}\ \bibnamefont {Liu}},\
  }\href@noop {} {\bibfield  {journal} {\bibinfo  {journal} {Journal of Sound
  and Vibration}\ }\textbf {\bibinfo {volume} {229}},\ \bibinfo {pages} {1171}
  (\bibinfo {year} {2000})}\BibitemShut {NoStop}%
\bibitem [{\citenamefont {Xia}(2003)}]{xia2003modelling}%
  \BibitemOpen
  \bibfield  {author} {\bibinfo {author} {\bibfnamefont {F.}~\bibnamefont
  {Xia}},\ }\href@noop {} {\bibfield  {journal} {\bibinfo  {journal} {Journal
  of Sound and Vibration}\ }\textbf {\bibinfo {volume} {265}},\ \bibinfo
  {pages} {1063} (\bibinfo {year} {2003})}\BibitemShut {NoStop}%
\bibitem [{\citenamefont {Korman}\ and\ \citenamefont
  {Li}(2014)}]{korman2014harmonic}%
  \BibitemOpen
  \bibfield  {author} {\bibinfo {author} {\bibfnamefont {P.}~\bibnamefont
  {Korman}}\ and\ \bibinfo {author} {\bibfnamefont {Y.}~\bibnamefont {Li}},\
  }\href@noop {} {\bibfield  {journal} {\bibinfo  {journal} {Acta Mathematica
  Scientia}\ }\textbf {\bibinfo {volume} {34}},\ \bibinfo {pages} {1025}
  (\bibinfo {year} {2014})}\BibitemShut {NoStop}%
\bibitem [{\citenamefont {Lima}\ and\ \citenamefont
  {Sampaio}(2015)}]{lima2015stick}%
  \BibitemOpen
  \bibfield  {author} {\bibinfo {author} {\bibfnamefont {R.}~\bibnamefont
  {Lima}}\ and\ \bibinfo {author} {\bibfnamefont {R.}~\bibnamefont {Sampaio}},\
  }\href@noop {} {\bibfield  {journal} {\bibinfo  {journal} {Journal of Sound
  and Vibration}\ }\textbf {\bibinfo {volume} {353}},\ \bibinfo {pages} {259}
  (\bibinfo {year} {2015})}\BibitemShut {NoStop}%
\bibitem [{\citenamefont {Binnig}\ \emph {et~al.}(1986)\citenamefont {Binnig},
  \citenamefont {Quate},\ and\ \citenamefont {Gerber}}]{binnig1986afm}%
  \BibitemOpen
  \bibfield  {author} {\bibinfo {author} {\bibfnamefont {G.}~\bibnamefont
  {Binnig}}, \bibinfo {author} {\bibfnamefont {C.~F.}\ \bibnamefont {Quate}}, \
  and\ \bibinfo {author} {\bibfnamefont {C.}~\bibnamefont {Gerber}},\ }\href
  {\doibase 10.1103/PhysRevLett.56.930} {\bibfield  {journal} {\bibinfo
  {journal} {Physical review letters}\ }\textbf {\bibinfo {volume} {56}},\
  \bibinfo {pages} {930} (\bibinfo {year} {1986})}\BibitemShut {NoStop}%
\bibitem [{\citenamefont {Mate}\ \emph {et~al.}(1987)\citenamefont {Mate},
  \citenamefont {McClelland}, \citenamefont {Erlandsson},\ and\ \citenamefont
  {Chiang}}]{mate1987atomic}%
  \BibitemOpen
  \bibfield  {author} {\bibinfo {author} {\bibfnamefont {C.~M.}\ \bibnamefont
  {Mate}}, \bibinfo {author} {\bibfnamefont {G.~M.}\ \bibnamefont
  {McClelland}}, \bibinfo {author} {\bibfnamefont {R.}~\bibnamefont
  {Erlandsson}}, \ and\ \bibinfo {author} {\bibfnamefont {S.}~\bibnamefont
  {Chiang}},\ }\bibfield  {booktitle} {\emph {\bibinfo {booktitle} {Scanning
  Tunneling Microscopy}},\ }\href {\doibase 10.1103/PhysRevLett.59.1942}
  {\bibfield  {journal} {\bibinfo  {journal} {Physical Review Letters}\
  }\textbf {\bibinfo {volume} {59}},\ \bibinfo {pages} {1942} (\bibinfo {year}
  {1987})}\BibitemShut {NoStop}%
\bibitem [{\citenamefont {Krim}\ and\ \citenamefont
  {Widom}(1988)}]{krim1988damping}%
  \BibitemOpen
  \bibfield  {author} {\bibinfo {author} {\bibfnamefont {J.}~\bibnamefont
  {Krim}}\ and\ \bibinfo {author} {\bibfnamefont {A.}~\bibnamefont {Widom}},\
  }\href {\doibase 10.1103/PhysRevB.38.12184} {\bibfield  {journal} {\bibinfo
  {journal} {Physical Review B}\ }\textbf {\bibinfo {volume} {38}},\ \bibinfo
  {pages} {12184} (\bibinfo {year} {1988})}\BibitemShut {NoStop}%
\bibitem [{\citenamefont {Krim}(2012)}]{krim2012friction}%
  \BibitemOpen
  \bibfield  {author} {\bibinfo {author} {\bibfnamefont {J.}~\bibnamefont
  {Krim}},\ }\href {\doibase 10.1080/00018732.2012.706401} {\bibfield
  {journal} {\bibinfo  {journal} {Advances in Physics}\ }\textbf {\bibinfo
  {volume} {61}},\ \bibinfo {pages} {155} (\bibinfo {year} {2012})}\BibitemShut
  {NoStop}%
\bibitem [{\citenamefont {Krim}\ \emph {et~al.}(1991)\citenamefont {Krim},
  \citenamefont {Solina},\ and\ \citenamefont
  {Chiarello}}]{krim1991nanotribology}%
  \BibitemOpen
  \bibfield  {author} {\bibinfo {author} {\bibfnamefont {J.}~\bibnamefont
  {Krim}}, \bibinfo {author} {\bibfnamefont {D.}~\bibnamefont {Solina}}, \ and\
  \bibinfo {author} {\bibfnamefont {R.}~\bibnamefont {Chiarello}},\ }\href
  {\doibase 10.1103/PhysRevLett.66.181} {\bibfield  {journal} {\bibinfo
  {journal} {Physical Review Letters}\ }\textbf {\bibinfo {volume} {66}},\
  \bibinfo {pages} {181} (\bibinfo {year} {1991})}\BibitemShut {NoStop}%
\bibitem [{\citenamefont {Yoshizawa}\ \emph {et~al.}(1993)\citenamefont
  {Yoshizawa}, \citenamefont {Chen},\ and\ \citenamefont
  {Israelachvili}}]{Yoshizawa1993fundamental}%
  \BibitemOpen
  \bibfield  {author} {\bibinfo {author} {\bibfnamefont {H.}~\bibnamefont
  {Yoshizawa}}, \bibinfo {author} {\bibfnamefont {Y.~L.}\ \bibnamefont {Chen}},
  \ and\ \bibinfo {author} {\bibfnamefont {J.}~\bibnamefont {Israelachvili}},\
  }\href {\doibase 10.1021/j100118a033} {\bibfield  {journal} {\bibinfo
  {journal} {The Journal of Physical Chemistry}\ }\textbf {\bibinfo {volume}
  {97}},\ \bibinfo {pages} {4128} (\bibinfo {year} {1993})}\BibitemShut
  {NoStop}%
\bibitem [{\citenamefont {Yoshizawa}\ and\ \citenamefont
  {Israelachvili}(1993)}]{yoshizawa1993fundamentaloi}%
  \BibitemOpen
  \bibfield  {author} {\bibinfo {author} {\bibfnamefont {H.}~\bibnamefont
  {Yoshizawa}}\ and\ \bibinfo {author} {\bibfnamefont {J.}~\bibnamefont
  {Israelachvili}},\ }\href {\doibase 10.1021/j100145a031} {\bibfield
  {journal} {\bibinfo  {journal} {The Journal of Physical Chemistry}\ }\textbf
  {\bibinfo {volume} {97}},\ \bibinfo {pages} {11300} (\bibinfo {year}
  {1993})}\BibitemShut {NoStop}%
\bibitem [{\citenamefont {Riedo}\ \emph {et~al.}(2003)\citenamefont {Riedo},
  \citenamefont {Gnecco}, \citenamefont {Bennewitz}, \citenamefont {Meyer},\
  and\ \citenamefont {Brune}}]{riedo2003interaction}%
  \BibitemOpen
  \bibfield  {author} {\bibinfo {author} {\bibfnamefont {E.}~\bibnamefont
  {Riedo}}, \bibinfo {author} {\bibfnamefont {E.}~\bibnamefont {Gnecco}},
  \bibinfo {author} {\bibfnamefont {R.}~\bibnamefont {Bennewitz}}, \bibinfo
  {author} {\bibfnamefont {E.}~\bibnamefont {Meyer}}, \ and\ \bibinfo {author}
  {\bibfnamefont {H.}~\bibnamefont {Brune}},\ }\href@noop {} {\bibfield
  {journal} {\bibinfo  {journal} {Physical review letters}\ }\textbf {\bibinfo
  {volume} {91}},\ \bibinfo {pages} {084502} (\bibinfo {year}
  {2003})}\BibitemShut {NoStop}%
\bibitem [{\citenamefont {Tambe}\ and\ \citenamefont
  {Bhushan}(2005)}]{tambe2005friction}%
  \BibitemOpen
  \bibfield  {author} {\bibinfo {author} {\bibfnamefont {N.~S.}\ \bibnamefont
  {Tambe}}\ and\ \bibinfo {author} {\bibfnamefont {B.}~\bibnamefont
  {Bhushan}},\ }\href {\doibase 10.1088/0957-4484/16/10/054} {\bibfield
  {journal} {\bibinfo  {journal} {Nanotechnology}\ }\textbf {\bibinfo {volume}
  {16}},\ \bibinfo {pages} {2309} (\bibinfo {year} {2005})}\BibitemShut
  {NoStop}%
\bibitem [{\citenamefont {Urbakh}\ and\ \citenamefont
  {Meyer}(2010)}]{urbakh2010nanotribology}%
  \BibitemOpen
  \bibfield  {author} {\bibinfo {author} {\bibfnamefont {M.}~\bibnamefont
  {Urbakh}}\ and\ \bibinfo {author} {\bibfnamefont {E.}~\bibnamefont {Meyer}},\
  }\href {\doibase 10.1038/nmat2599} {\bibfield  {journal} {\bibinfo  {journal}
  {Nature materials}\ }\textbf {\bibinfo {volume} {9}},\ \bibinfo {pages} {8}
  (\bibinfo {year} {2010})}\BibitemShut {NoStop}%
\bibitem [{\citenamefont {Chiu}\ \emph {et~al.}(2012)\citenamefont {Chiu},
  \citenamefont {Dogan}, \citenamefont {Volkmann}, \citenamefont {Klinke},\
  and\ \citenamefont {Riedo}}]{chiu2012adhesion}%
  \BibitemOpen
  \bibfield  {author} {\bibinfo {author} {\bibfnamefont {H.-C.}\ \bibnamefont
  {Chiu}}, \bibinfo {author} {\bibfnamefont {S.}~\bibnamefont {Dogan}},
  \bibinfo {author} {\bibfnamefont {M.}~\bibnamefont {Volkmann}}, \bibinfo
  {author} {\bibfnamefont {C.}~\bibnamefont {Klinke}}, \ and\ \bibinfo {author}
  {\bibfnamefont {E.}~\bibnamefont {Riedo}},\ }\href {\doibase
  10.1088/0957-4484/23/45/455706} {\bibfield  {journal} {\bibinfo  {journal}
  {Nanotechnology}\ }\textbf {\bibinfo {volume} {23}},\ \bibinfo {pages}
  {455706} (\bibinfo {year} {2012})}\BibitemShut {NoStop}%
\bibitem [{\citenamefont {Borovsky}\ \emph {et~al.}(2001)\citenamefont
  {Borovsky}, \citenamefont {Krim}, \citenamefont {Asif},\ and\ \citenamefont
  {Wahl}}]{borovsky2001measuring}%
  \BibitemOpen
  \bibfield  {author} {\bibinfo {author} {\bibfnamefont {B.}~\bibnamefont
  {Borovsky}}, \bibinfo {author} {\bibfnamefont {J.}~\bibnamefont {Krim}},
  \bibinfo {author} {\bibfnamefont {S.~S.}\ \bibnamefont {Asif}}, \ and\
  \bibinfo {author} {\bibfnamefont {K.}~\bibnamefont {Wahl}},\ }\href {\doibase
  10.1063/1.1413493} {\bibfield  {journal} {\bibinfo  {journal} {Journal of
  Applied Physics}\ }\textbf {\bibinfo {volume} {90}},\ \bibinfo {pages} {6391}
  (\bibinfo {year} {2001})}\BibitemShut {NoStop}%
\bibitem [{\citenamefont {Berg}\ and\ \citenamefont
  {Johannsmann}(2003)}]{berg2003high}%
  \BibitemOpen
  \bibfield  {author} {\bibinfo {author} {\bibfnamefont {S.}~\bibnamefont
  {Berg}}\ and\ \bibinfo {author} {\bibfnamefont {D.}~\bibnamefont
  {Johannsmann}},\ }\href {\doibase 10.1103/PhysRevLett.91.145505} {\bibfield
  {journal} {\bibinfo  {journal} {Physical review letters}\ }\textbf {\bibinfo
  {volume} {91}},\ \bibinfo {pages} {145505} (\bibinfo {year}
  {2003})}\BibitemShut {NoStop}%
\bibitem [{\citenamefont {Inoue}\ \emph {et~al.}(2012)\citenamefont {Inoue},
  \citenamefont {Machida}, \citenamefont {Taniguchi}, \citenamefont {Suzuki},
  \citenamefont {Ishikawa},\ and\ \citenamefont {Miura}}]{inoue2012dynamical}%
  \BibitemOpen
  \bibfield  {author} {\bibinfo {author} {\bibfnamefont {D.}~\bibnamefont
  {Inoue}}, \bibinfo {author} {\bibfnamefont {S.}~\bibnamefont {Machida}},
  \bibinfo {author} {\bibfnamefont {J.}~\bibnamefont {Taniguchi}}, \bibinfo
  {author} {\bibfnamefont {M.}~\bibnamefont {Suzuki}}, \bibinfo {author}
  {\bibfnamefont {M.}~\bibnamefont {Ishikawa}}, \ and\ \bibinfo {author}
  {\bibfnamefont {K.}~\bibnamefont {Miura}},\ }\href {\doibase
  10.1103/PhysRevB.86.115411} {\bibfield  {journal} {\bibinfo  {journal}
  {Physical Review B}\ }\textbf {\bibinfo {volume} {86}},\ \bibinfo {pages}
  {115411} (\bibinfo {year} {2012})}\BibitemShut {NoStop}%
\bibitem [{\citenamefont {Nigues}\ \emph {et~al.}(2014)\citenamefont {Nigues},
  \citenamefont {Siria}, \citenamefont {Vincent}, \citenamefont {Poncharal},\
  and\ \citenamefont {Bocquet}}]{nigues2014ultrahigh}%
  \BibitemOpen
  \bibfield  {author} {\bibinfo {author} {\bibfnamefont {A.}~\bibnamefont
  {Nigues}}, \bibinfo {author} {\bibfnamefont {A.}~\bibnamefont {Siria}},
  \bibinfo {author} {\bibfnamefont {P.}~\bibnamefont {Vincent}}, \bibinfo
  {author} {\bibfnamefont {P.}~\bibnamefont {Poncharal}}, \ and\ \bibinfo
  {author} {\bibfnamefont {L.}~\bibnamefont {Bocquet}},\ }\href {\doibase
  10.1038/nmat3985} {\bibfield  {journal} {\bibinfo  {journal} {Nature
  materials}\ }\textbf {\bibinfo {volume} {13}},\ \bibinfo {pages} {688}
  (\bibinfo {year} {2014})}\BibitemShut {NoStop}%
\bibitem [{\citenamefont {Berg}\ \emph {et~al.}(2003)\citenamefont {Berg},
  \citenamefont {Prellberg},\ and\ \citenamefont
  {Johannsmann}}]{berg2003nonlinear}%
  \BibitemOpen
  \bibfield  {author} {\bibinfo {author} {\bibfnamefont {S.}~\bibnamefont
  {Berg}}, \bibinfo {author} {\bibfnamefont {T.}~\bibnamefont {Prellberg}}, \
  and\ \bibinfo {author} {\bibfnamefont {D.}~\bibnamefont {Johannsmann}},\
  }\href {\doibase 10.1063/1.1523647} {\bibfield  {journal} {\bibinfo
  {journal} {Review of scientific instruments}\ }\textbf {\bibinfo {volume}
  {74}},\ \bibinfo {pages} {118} (\bibinfo {year} {2003})}\BibitemShut
  {NoStop}%
\end{thebibliography}
\end{document}